\documentclass[aps,reprint,amsmath,amssymb,showpacs,prx,floatfix]{revtex4-2}
\usepackage{amsmath}
\usepackage{amsfonts}
\usepackage{amssymb}
\usepackage[dvipsnames]{xcolor}
\usepackage{graphicx,color,subeqnarray}
\usepackage{dcolumn}
\usepackage{multirow}
\usepackage{bm}

\usepackage{verbatim}

\begin{document}

\title{Transition path time over a barrier of a colloidal particle in a viscoelastic bath
}
\author{Brandon
\surname{R. Ferrer}}
\affiliation{Instituto de F\'isica, Universidad Nacional Aut\'onoma de M\'exico,\\ Cd. de M\'exico, C.P. 04510, Mexico}

\author{Alejandro V.  
\surname{Arzola}}
\affiliation{Instituto de F\'isica, Universidad Nacional Aut\'onoma de M\'exico,\\ Cd. de M\'exico, C.P. 04510, Mexico}

\author{Denis \surname{Boyer}}
\affiliation{Instituto de F\'isica, Universidad Nacional Aut\'onoma de M\'exico,\\ Cd. de M\'exico, C.P. 04510, Mexico}

\author{Juan Ruben \surname{Gomez-Solano}}
\email[]{r\_gomez@fisica.unam.mx}
\affiliation{Instituto de F\'isica, Universidad Nacional Aut\'onoma de M\'exico,\\ Cd. de M\'exico, C.P. 04510, Mexico}

\date{\today}

\begin{abstract}
We experimentally study the statistics of the transition path time taken by a submicron bead to successfully traverse an energy barrier created by two optical tweezers in two prototypical viscoelastic fluids, namely, aqueous polymer and micellar solutions. We find a very good agreement between our experimental distributions and a theoretical  expression derived from the generalized Langevin equation for the particle motion. Our results reveal that the mean transition path time measured in such viscoelastic fluids have a non-trivial dependence on the barrier curvature and they can be significantly reduced when compared with those determined in Newtonian fluids of the same zero-shear viscosity. We verify that the decrease of the mean transition path time can be described in terms of an effective viscosity that quantitatively coincides with that measured by linear microrheology at a frequency determined by the reactive mode that gives rise to the unstable motion over the barrier. Therefore, our results uncover the linear response of the particle during its thermally activated escape from a metastable state even when taking place in a non-Markovian bath.

\end{abstract}

\maketitle
\section{Introduction}

Investigating the motion of mesoscopic objects through complex fluid environments is of prime importance in many  processes such as the walk of molecular motors \cite{haenggi2009,hoffmann2016}, intracellular transport  \cite{bressloff2013,mogre2020}, nanofriction \cite{vanossi2012,reichhardt2017}, relaxations of colloidal suspensions \cite{schweizer2003,ma2019}, etc. 
In these systems, the particles must be assisted by sufficiently strong forces to escape from the local minima of their energy landscape in order to explore their surroundings. Over the past few decades, these types of small-scaled processes have motivated the experimental study of transport phenomena of colloidal particles dispersed in fluids across multistable energy landscapes carefully crafted by means of optical techniques. Among others, these include: diffusion over periodic and random potentials \cite{lee2006,siler2010,dalle2011,hanes2012,hanes2013,barakat2023,nagella2023}, synchronization effects \cite{juniper2015,juniper2017,abbott2019}, collective motion due to hydrodynamic interactions \cite{lutz2006,lips2022}, fluctuation-dissipation relations far from equilibrium \cite{blickle2007,gomezsolano2009,gomezsolano2011}, non-equilibrium phase transitions \cite{bohlein2012,zaidouny2013,
gomezsolano2015}, and directed motion through tilted potentials
\cite{roichmann2007,evstigneev2008,juniper2016,abbott_2019}. In these situations, one is often interested in the time that a particle spends to travel over a characteristic length-scale of the optical potential. A fundamental time-scale that has been a major subject of interest since the development of Kramers' theory is the mean time taken by a particle to escape over a barrier from a potential well~\cite{kramers1940}, where the necessary energy is supplied by the  molecules of the medium. Indeed, a large body of theoretical research has focused on calculating the mean escape times of Brownian particles through thermal activation in fluids of varying physical properties, including those exhibiting long-time memory~\cite{hanggi1982,carmeli1984,munakata1985,talkner1988,goychuk2009,goychuk2012}, low friction \cite{carmeli1983,melnikov1986,pollak1989,kappler2018}, shear flow \cite{kienle2011}, hydrodynamic backflow \cite{cherayil2022}, and particle interactions \cite{kumar2024,kumar_2024}. Some predictions on the Arrhenius-like properties of the mean escape times have been confirmed in experiments using micron-sized beads hopping across bistable optical potentials with sufficiently high barriers in viscous liquids \cite{simon1992,mccann1999,wu2009,zijlstra2020,ferrer2024}, gases \cite{rondin2017}, and viscoelastic fluids \cite{ferrer2021,ginot2022}. 

A second important time-scale that characterizes the thermally assisted  motion of a colloidal particle across a multistable energy landscape is the mean duration of its transition paths. A transition path is a trajectory segment that traverses the potential barrier from one specific position to another on the opposite side, without recrossing none of these two positions. Therefore, the duration of a transition path is significantly shorter than the corresponding escape time, since the latter is a first passage time comprising all unsuccessful attempts to surmount the barrier before an actual transition path takes place. Even though transitions paths encode valuable information on the mechanisms that enable particle transport, their study has been mainly conducted in the context of conformational changes  of macromolecules~\cite{hummer2004,zhang2007,chaudhury2010,malinin2010,kim2015,makarov2015,pollak2016,laleman2017,carlon2018,berezhkovskii2018,pyo2018,medina2018,
berezhkovskii2019,caraglio2020,li2020,singh2021,singh_2021,dutta2022}, e.g. in protein and nucleic acids folding experiments \cite{chung2009,neupane2012,chung2012,lannon2013,chung2014,chung2015,neupane2018,kim2020,tripathi2022}. Indeed, there are only a few experimental works that have analyzed the statistics of transition paths of colloidal beads hopping in bistable potentials in viscous (Newtonian) liquids~\cite{zijlstra2020,ferrer2024,lyons2024}, which exhibit properties described by diffusive models. Nevertheless, many soft materials display viscoelastic behavior due to their macromolecular microstructure~\cite{larson1999}, thus leading to memory effects on the dynamics of Brownian particles dispersed in them~\cite{squires2010}. Since many transport phenomena commonly take place in viscoelastic environments \cite{caspi2002,weber2010,tabei2013,ahmed2018,zhou2020,
burghelea2020,gaojin2021}, it is of prime importance to examine the properties of transition paths of colloidal particles under these conditions.

\begin{figure*}[ht!]
\centering
\includegraphics[width=0.95\textwidth]{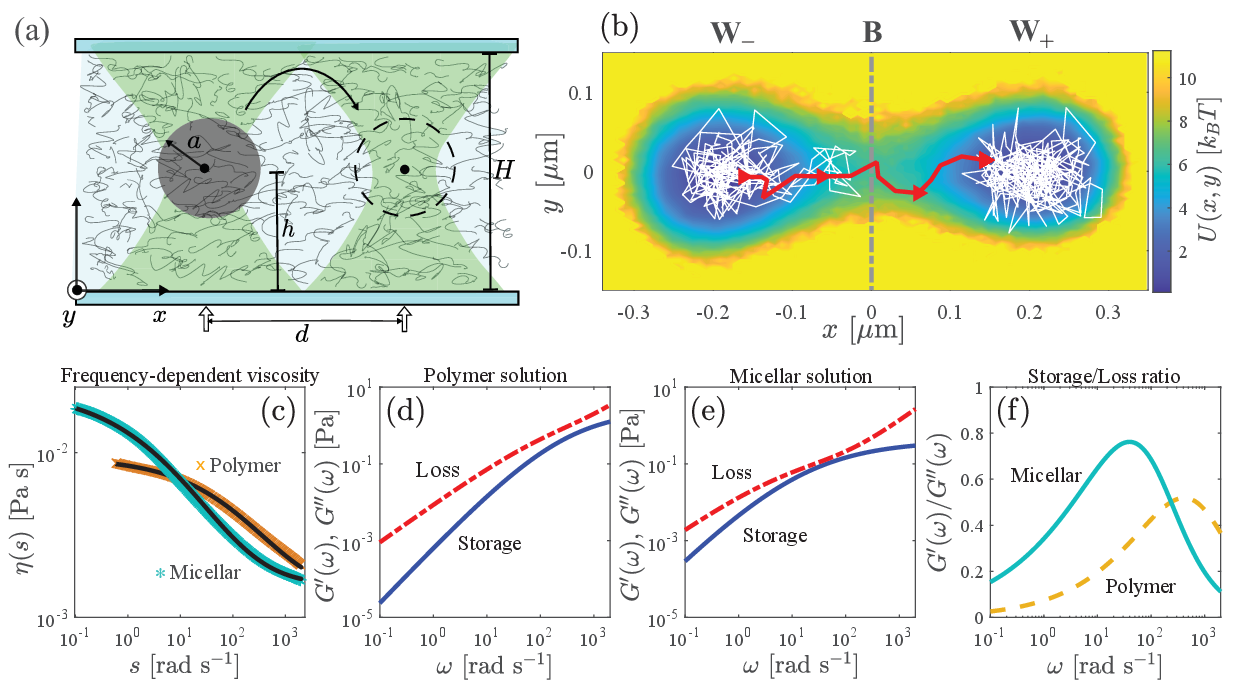}
 \caption{(a) Sketch of a bistable potential created by means of two optical tweezers separated by a distance $d$, trapping a colloidal bead of radius $a$ in a fluid at a height $h$ from the lower wall of a sample cell of thickness $H$. (b) Color map of a double-well potential acting on a bead in the transverse plane $xy$. Two portions of a same trajectory where the particle dwells in each potential well $\mathbf{W}_{\pm}$ are depicted in white, while the segment representing a successful crossing event from $\mathbf{W}_-$ to $\mathbf{W}_+$ over the barrier $\mathbf{B}$ is represented in red.
(c) Frequency-dependent viscosities of the polymer ($\times$, in orange) and micellar solutions ($*$, in turquoise), experimentally determined by passive microrheology. Solid lines represent the best nonlinear fit of the experimental data  to Eq.~(\ref{eq:etas}). The fitting parameters are: $\eta_0=0.0093$ Pa s, $\eta_\infty=0.0014$ Pa s, $\tau_0=0.003$ s, $\alpha=0.471$ (polymer solution), and $\eta_0=0.023$ Pa s, $\eta_\infty=0.0013$ Pa s, $\tau_0=0.0562$ s, $\alpha=0.405$ (micellar solution). (d-e) Storage (solid line) and loss modulus (dotted-dashed line) of the polymer and micellar solution, respectively, determined from the fitting curves of $\eta(s)$. (f) Ratio of the storage to the loss modulus as a function of frequency, for the polymer (dashed line) and micellar solution (solid line).}
\label{fig:Figure1}
\end{figure*}

In this article we experimentally study the statistical properties of the transition path time of a colloidal bead crossing an energy barrier separating two optical potential wells generated by two optical tweezers. Our goal is to elucidate the role of the viscoelasticity of the medium on the particle motion over a barrier whose shape can be approximated as an inverted parabola. The experiments are performed using two types of complex fluids, namely, a polymer and a micellar aqueous solution, both exhibiting viscoelasticity. Our results are compared with those derived from a theoretical model based on the generalizad Langevin equation including the effect of viscoelasticity through a friction memory kernel combined with a colored thermal noise acting on the particle. Based on our experimental findings, we provide an effective description of the observed effects that reveal the role of the frequency-dependence of the viscoelastic properties of the medium on the barrier-crossing process of the particle.

\section{Experimental description}\label{sect:exp}
We prepared a highly diluted dispersion of spherical silica beads (diameter $2a=0.5~ \mu\mathrm{m}$) in the fluid of interest, which was confined in a hermetic cell made of a microscope slide adhered to a coverslip by a double-sided adhesive tape (separation $H \approx 100~\mu$m) and sealed with epoxy glue. Then, a double-well optical potential was generated by focusing two Gaussian laser beams (wavelength 532~nm) inside the sample cell by means of an oil-immersion objective (100$\times$, $\mathrm{NA} = 1.3$). As depicted in Fig. \ref{fig:Figure1}(a), as a result of the optical forces, a single bead was trapped at room temperature $T=22~^{\circ}$C at a distance of $h \approx 10\, \mu\mathrm{m}$ above the bottom wall of the cell, thereby avoiding hydrodynamic interactions with the cell walls or with other particles. The two optical traps were separated by a distance of $d\approx 0.5~\mu$m, allowing the particle to hop between the two corresponding potential wells by thermal activation, thus effectively behaving as a bistable potential. Videos of the particle moving in such an energy landscape were recorded using a CMOS
camera at a sampling frequency of $f_s = 2500$~Hz for 1 hour. From these videos, the coordinates $(x,y)$ of the center of mass of the bead projected on the transverse plane to the beam direction were measured by standard particle-tracking methods with a spatial resolution of 5~nm. The two-dimensional profile of the optical potential was reconstructed from these coordinates by means of the Boltzmann distribution
\begin{equation}\label{eq:Uprof2D}
    \rho_{eq}(x,y) = \rho_0 \exp \left[ - \frac{U(x,y)}{k_B T}\right],
\end{equation}
where $\rho_0$ is a normalization constant and $k_B = 1.380649 \times 10^{-23}$~J~K$^{-1}$ is the Boltzmann constant. As shown in Fig. \ref{fig:Figure1}(b), this potential exhibits two wells ($\textbf{W}_-$ and $\textbf{W}_+$), separated by an energy barrier ($\textbf{B}$), whose height depends on the laser intensity of each optical trap. The straight line connecting the minima of $\textbf{W}_{\pm}$ and the maximum of $\textbf{B}$ defines the $x$-axis ($y=0$), whereas the $y$-axis ($x = 0$) corresponds to the perpendicular direction, where the origin $(0,0)$ is chosen as the location of the maximum of $\textbf{B}$, as shown in Fig. \ref{fig:Figure1}(b). The detailed shape of $U(x,y)$ depends on the specific particle and can be modified by slightly changing the laser power or the separation between the two beams. Therefore, the parameters characterizing the potential landscape must be determined for each trapped particle.

Two complex fluids with distinct types of macromolecules were used as non-Markovian baths for the trapped colloidal beads. The first one was a polymer solution of polyethylene oxide (molecular weight $4\times 10^6$~Da) mixed in ultrapure water at 0.1\%~wt, a concentration at which the polymer chains were non-entangled but inter-chain interactions were not negligible \cite{ebagninin2009}. The second one consisted of a micellar solution composed of cetylpyridinium chloride and sodium salicylate at equimolar concentration of 5~mM in ultrapure water, at which the surfactant molecules self-assembled in worm-like micelles \cite{rehage88}.
Both solutions became homogeneous and transparent by continuous stirring over 24 hours, and exhibited viscoelastic response due to the presence of the suspended macromolecules. The viscoelasticity of both solutions was experimentally characterized by means of passive microrheology using a single optical trap~\cite{darabi2023}. In both cases, the viscoelasticity can be described by the following fitting function for the frequency-dependent viscosity
\begin{equation}\label{eq:etas}
    \eta(s) = \eta_{\infty} + \frac{\eta_0 - \eta_{\infty}}{\left[ 1 + (\tau_0 s)^{\alpha} \right]^{1/\alpha}},
\end{equation}
where $\eta(s)$ is the Laplace transform of the stress relaxation modulus, $G(t)$, i.e., $\eta(s) = \int_0^{\infty} dt \, e^{-st} G(t)$ \cite{larson1999}. In Eq. (\ref{eq:etas}), $s$ is a complex-valued frequency, $\eta_0$ and $\eta_{\infty}$ are the fluid's zero-shear viscosity and the solvent viscosity, respectively, $\tau_0$ is a stress relaxation time, and $0 < \alpha < 1$ is an exponent that quantifies the deviation of the viscoelasticity from the ideal Newtonian ($\alpha = 0$) and Maxwellian ($\alpha = 1$) behaviors. The function in Eq. (\ref{eq:etas}) captures the linear rheological properties of the fluid in a rather compact form, which is convenient for an easier comparison of the experimental results with the theoretical model of particle motion of the following section. The dependence of $\eta(s)$ for a real positive $s$ is shown in Fig. \ref{fig:Figure1}(c), where the micellar solution exhibits a more pronounced decrease with frequency from $\eta(s \rightarrow 0) = \eta_0$ to $\eta(s \rightarrow \infty) = \eta_{\infty}$ than the polymer solution, which can be attributed to the larger degree of elasticity of the former. 
Indeed, from the fitting function for $\eta(s)$ of Eq. (\ref{eq:etas}), one can obtain the complex shear modulus for each fluid, $G^*(\omega) = G'(\omega) + iG''(\omega)$, by means of the relation $G^*(\omega) = i\omega \eta(s = i\omega)= i\omega \int_{-\infty}^{\infty} dt \, e^{-i\omega t} G(t)$. As displayed in Figs. \ref{fig:Figure1}(d) and \ref{fig:Figure1}(e), both fluids exhibit non-trivial frequency dependent storage, $G'(\omega)$, and loss modulus, $G''(\omega)$, whose ratio $G'(\omega)/G''(\omega)$ confirms that elastic effects are more pronounced for the micellar than for the polymer solution, as shown in Fig. \ref{fig:Figure1}(f) 

Additionally, control experiments of beads in bistable optical potentials were conducted in Newtonian mixtures of glycerol at 55\%~wt and 70\%~wt in water. At such concentrations, the frequency-independent viscosities of these mixtures are $0.0076$~Pa~s and $0.021$~Pa~s, respectively, and comparable to the values of the zero-shear viscosities of the polymer fluid ($\eta_0 = 0.0093$~Pa~s) and of the micellar one ($\eta_0 = 0.023$~Pa~s).

\section{Transition path time distribution}\label{sect:TPTD}

\begin{figure*}
\centering
 \includegraphics[width=0.95\textwidth]{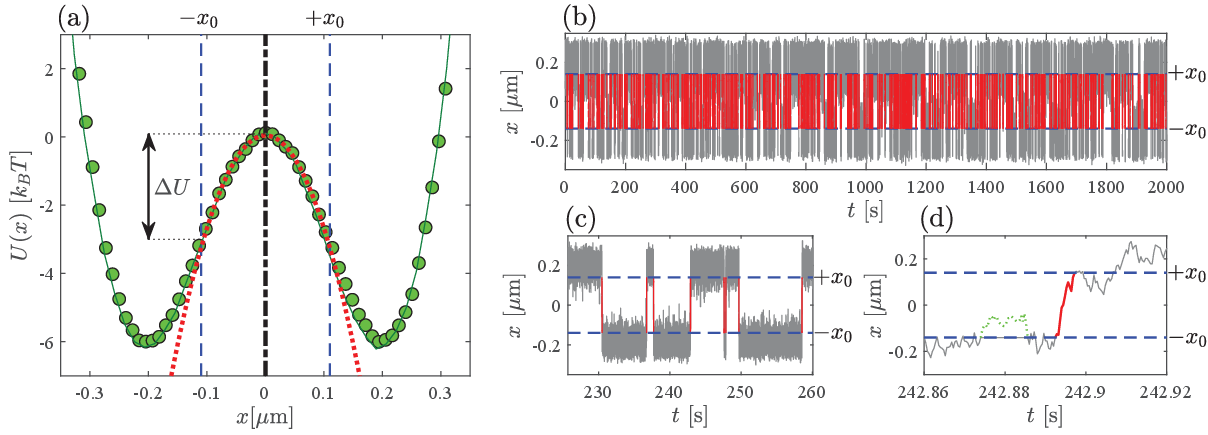}
 \caption{(a) One-dimensional potential $U(x)$ determined from the equilibrium distribution of $x$ by means of Eq. (\ref{eq:Uprof1D}) ($\circ$ symbol), or by evaluating $U(x,y)$ at $y=0$ (solid line). Vertical dashed lines indicate the boundaries of a transition region $|x|\le x_0$ centered around the barrier top at $x=0$. Dotted red lines represent the quadratic fit given by Eq. (\ref{eq:invparab}) of the barrier with height $\Delta U$ measured relative to the boundaries at $\pm x_0$. (b) Example of a bead trajectory, $x(t)$, over 2000~s, showing the stochastic hopping between the potential wells. The transition paths through the region with $x_0 = 150$~nm (dashed horizontal lines) are depicted in red. (c) Expanded view of Fig.~\ref{fig:Figure2}(b) over 35~s illustrating in red transitions paths from $-x_0$ to $+x_0$ and from $+x_0$ to $-x_0$. (d) Expanded view of Fig.~\ref{fig:Figure2}(c) over 60~ms, exemplifying in red an actual transition path from $-x_0$ to $+x_0$ and a segment of the trajectory inside the transition region that does not correspond to a transition path (dotted green line).}
\label{fig:Figure2}
\end{figure*}

Due to the symmetry of the potential $U(x,y)$ with respect to the $x$-axis, we limit the analysis of the transition paths of the bead
to this coordinate. To this end, from the marginal probability density function of $x$, $\rho_{eq}(x) = \int_{-\infty}^{\infty}dy \, \rho_{eq}(x,y)$, we compute the one-dimensional potential of mean force
\begin{equation}\label{eq:Uprof1D}
    U(x) = -k_B T \ln \rho_{eq}(x) + u_0,
\end{equation}
where $u_0$ is an arbitrary constant. An example of such a potential is displayed in Fig. \ref{fig:Figure2}(a). We notice that $U(x)$ coincides with $U(x, y=0)$, thereby showing that the thermally activated motion of the bead can be effectively described as a one-dimensional problem. In the same figure, we also define the transition region $| x | \le x_0$, which is defined by the two boundaries $x = -x_0$ and $x = +x_0$ symmetrically located around $x = 0$, where $x_0$  is a positive threshold parameter. The values of $x_0$ are selected such that $U(x)$ can be accurately fitted by an inverted parabola in the interval $-x_0 \le x \le +x_0$, namely,
\begin{equation}\label{eq:invparab}
    U(x) = U(0) - \frac{1}{2} \kappa x^2,
\end{equation}
with curvature $ -\kappa = U''(0) < 0$, thus avoiding the non-linear effects due to the potential wells $\textbf{W}_{\pm}$. The barrier height of the transition region is defined as $\Delta U \equiv U(0) - U(x_0) = \frac{1}{2} \kappa x_0^2$ and it is chosen to be at least $2.5 k_B T$. For such values of $\Delta U$ the barrier is sufficiently high, allowing to derive an analytical solution as shown below. With these restrictions, typical values of $x_0$ and $\kappa$ range from 90~nm to 180 nm and from 1.57 $\times$ 10$^{-6}$ N~m$^{-1}$ to 2.92 $\times$ 10$^{-6}$ N~m$^{-1} $, respectively.

\begin{figure*}
\centering
\includegraphics[width=0.965\textwidth]{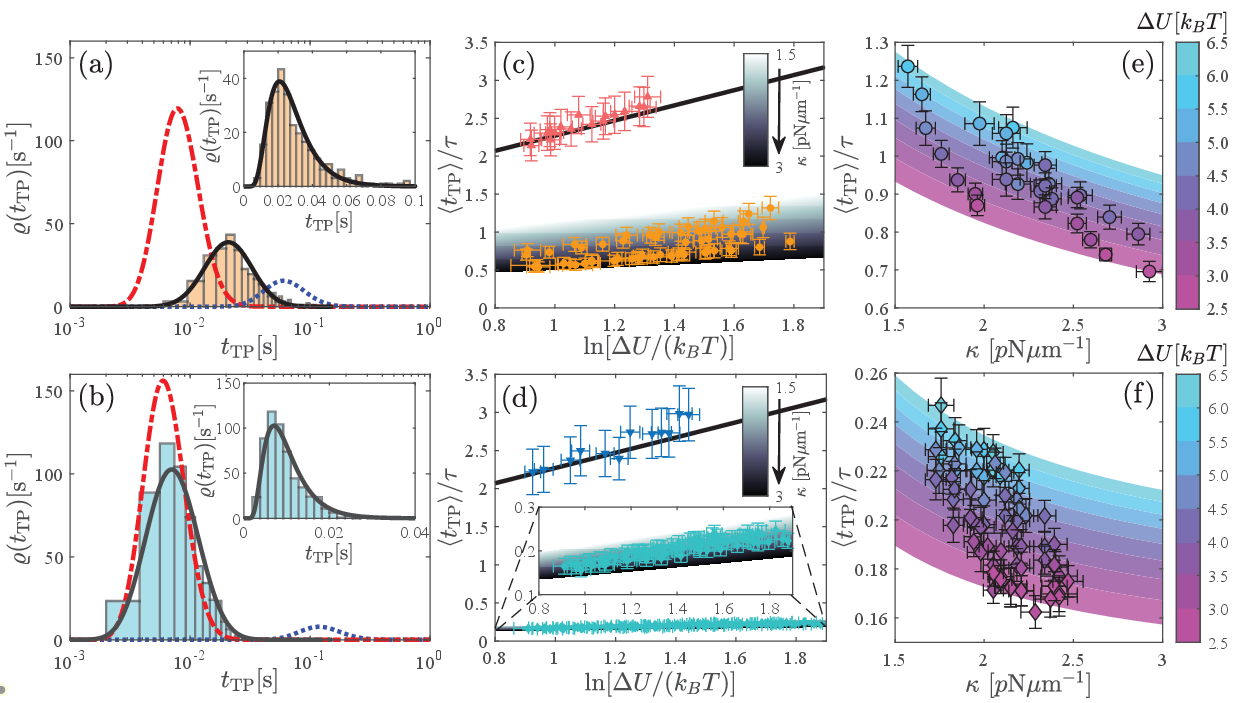}
 \caption{Semilog plots of the experimental transition path time distributions (vertical bars) of particles moving 
 over the barriers of similar optical potentials in the two viscoelastic fluids: (a) polymer solution, with $\kappa = 1.67~\mathrm{pN}\mu \mathrm{m}^{-1}$, $x_0 =0.15~\mu\mathrm{m}$, $\Delta U=4.63 k_B T$; (b) micellar solution, with $\kappa = 1.74~\mathrm{pN}\mu \mathrm{m}^{-1}$, $x_0 =0.15~\mu\mathrm{m}$, $\Delta U=4.81 k_B T$. Colored dotted and dotted-dashed lines correspond to the Newtonian model in Eq. (\ref{eq:rhotpt_newton}) with the corresponding values of $\eta_0$ and $\eta_{\infty}$, respectively. Solid lines are obtained from the viscoelastic model in Eq. (\ref{eq:rhotpt_viscoelastic}). Insets in (a) and (b) represent the same distributions in linear scale. (c-d) Mean transition path time rescaled by $\tau\equiv 6\pi a \eta_0/\kappa$ as a function of the barrier height measured for various potentials in: (c) the polymer solution (orange $\circ$ symbols), the glycerol-water mixture at 55\%~wt (red $\triangle$ symbols);
 (d) the micellar solution (turquoise $\diamond$ symbols), and the glycerol-water mixture at 70\%~wt (blue $\bigtriangledown$ symbols).
 In (c) and (d), the shaded area is determined using Eq. (\ref{eq:rhotpt_viscoelastic}), where the values of $\kappa$ are indicated by the grayscale, whereas the solid line in the upper part corresponds to the viscous case in Eq. (\ref{eq:MTPTNewtonian}). The inset in (d) is a zoom of the results for the micellar solution. (e-f)
 Mean transition path time rescaled by $\tau$ as a function of the barrier curvature in: (e) the polymer solution ($\circ$ symbols), and (f) the micellar solution ($\diamond$ symbols).
 In (e) and (f), each stripe is obtained from evaluating numerically the first moment of the distribution (\ref{eq:rhotpt_viscoelastic}) for a given barrier height with an increment of $0.5k_B T$. The color code for both the experimental and the theoretical results depicts the value of the barrier height.}
\label{fig:Figure3}
\end{figure*}

Once the transition region and its corresponding curvature  have been determined for a given potential, we measure the duration $t_{\mathrm{TP}}$ of many transition paths along a trajectory followed by the bead, as exemplified in Fig. \ref{fig:Figure2}(b).  By definition, a transition path corresponds to a portion of the trajectory that begins by entering the transition region at, say,
$x = -x_0$ and ends when reaching $x = +x_0$ for the first time without having returned to $-x_0$, like the segments shown in red in Figs~\ref{fig:Figure2}(b)-(d). Trajectory segments like the one depicted 
by a dotted green line in Figs.~\ref{fig:Figure2}(d), that enters the transition region at $-x_0$ and crosses this position again before reaching the boundary at $+x_0$, must be excluded from the analysis. Because of the symmetry of the potential $U(x)$ inside the transition region, for given values of $x_0$ and $\kappa$ the duration of the transition paths in the reverse direction, i.e. from $+x_0$ to $-x_0$, are also counted in the statistics of the corresponding $t_{TP}$, as illustrated in Figs.~\ref{fig:Figure2}(b) and \ref{fig:Figure2}(c). From $n$ experimental values of $t_{\mathrm{TP}}$ ($n=600$, typically) we determine the probability distribution of the transition path time, $\varrho(t_{\mathrm{TP}})$.
Figs. \ref{fig:Figure3}(a) and \ref{fig:Figure3}(b) display examples of $\varrho(t_{\mathrm{TP}})$ measured in the polymer and the micellar fluids, respectively, for similar energy barriers. As shown clearly in the insets, in both cases $\varrho(t_{\mathrm{TP}})$ exhibits a non-monotonic asymmetric shape around a single peak and vanishes as $t_{\mathrm{TP}} \rightarrow 0$ and $t_{\mathrm{TP}} \rightarrow \infty$. Notice that the shape of  $\varrho(t_{\mathrm{TP}})$ significantly differs from that of the probability distribution of the escape times in the Kramers' problem, which is an exponentially decaying function due to the Poissonian nature of the process for high barriers.

To gain some insight into the properties of these transition path time distributions, we first compare the experimental curves with those obtained from a purely diffusive model for the particle motion. In that Markovian case, for sufficiently high barriers a good analytic approximation of the transition path time distribution is given by \cite{zhang2007,laleman2017}
\begin{equation}\label{eq:rhotpt_newton}      \varrho(t_{\mathrm{TP}}) = \frac{1}{2\tau} \frac{\sqrt{\frac{\kappa x_0^2}{\pi k_B T}}}{\mathrm{erfc}\left( \frac{\kappa x_0^2}{2k_B T}\right)} \frac{\exp\left[ - \frac{\kappa x_0^2}{2k_B T} \coth \left( \frac{t_{\mathrm{TP}}}{2\tau}\right)\right]}{\sinh \left( \frac{t_{\mathrm{TP}}}{2\tau}\right) \sqrt{\sinh \left( \frac{t_{\mathrm{TP}}}{\tau}\right)}},
\end{equation}
where $\mathrm{erfc}(z) = \frac{2}{\sqrt{\pi}}\int_z^{\infty} dt \, e^{-t^2}$ is the complementary error function. In Eq. (\ref{eq:rhotpt_newton}), $\tau = \gamma/\kappa$ is the characteristic time-scale of the particle moving over the barrier of curvature $-\kappa$, where $\gamma = 6\pi a \eta$ is the friction coefficient of a solid sphere in a Newtonian fluid with constant viscosity $\eta$. In Figs. \ref{fig:Figure3}(a) and \ref{fig:Figure3}(b) we show the transition path time distributions computed from Eq. (\ref{eq:rhotpt_newton}) for the polymer and micellar solutions, respectively, using for each fluid the two characteristic values of the frequency-dependent viscosity given by Eq. (\ref{eq:etas}), i.e., $\eta_0$ and $\eta_{\infty}$. None of these curves describe accurately the experimental distributions, thereby revealing that the viscoelasticity of the medium strongly affects the statistical properties of the transition path time. Therefore, to go one step further, we model the non-Markovian particle dynamics over the barrier by making use of the generalized Langevin equation in the overdamped regime~\cite{zwanzig1973} 
\begin{equation}\label{eq:GLE}
    0 = - \int_0^t dt' \, \Gamma(t-t') \dot{x}(t') + \kappa x(t) + \zeta(t).
\end{equation}
In Eq. (\ref{eq:GLE}), $\Gamma(t-t')$ is the friction memory kernel, whose Laplace transform can be approximately expressed in the overdamped limit as $\tilde{\Gamma}(s) = 6\pi a \eta(s)$~\cite{darabi2023} with $\dot{x}(t')$ the particle velocity at time $t'$, while the inverted parabolic potential is extended to the whole domain $-\infty<x<\infty$ for convenience (see below and Appendix \ref{app:derivTPTP}), and $\zeta(t)$ is a colored Gaussian noise that satisfies
\begin{eqnarray}\label{eq:noise}
    \langle \zeta(t) \rangle & = & 0,\nonumber\\
   \langle \zeta(t) \zeta(0) \rangle & = & k_B T \Gamma(t).
\end{eqnarray}
As detailed in Appendix \ref{app:derivTPTP}, taking advantage of the linearity of Eq. (\ref{eq:GLE}) one can calculate the conditional probability density of the position $x$ at time $t > 0$ provided that the particle started from $-x_0$ at $t = 0$. This propagator is the solution of an effective Smoluchowski equation associated to Eq. (\ref{eq:GLE}). For high barriers $\Delta U$ (where the effects of re-crossing at the positions $\pm x_0$ can be readily neglected) one finds the following explicit expression for the transition path time distribution \cite{carlon2018,medina2018,singh2021} 
\begin{widetext}
\begin{equation}\label{eq:rhotpt_viscoelastic}\varrho(t_{\mathrm{TP}}) = \frac{1}{\mathrm{erfc}\left( \sqrt{\frac{\kappa x_0^2}{2 k_B T}}\right)}  \frac{\frac{d\chi(t_{\mathrm{TP}})}{dt_{\mathrm{TP}}}}{\chi(t_{\mathrm{TP}})-1} \sqrt{\frac{2 \kappa x_0^2}{\pi k_B T \left[ \chi(t_{\mathrm{TP}})^2 -1 \right]}} \exp \left\{ -\frac{\kappa x_0^2}{2 k_B T} \frac{\chi(t_{\mathrm{TP}}) + 1}{\chi(t_{\mathrm{TP}})-1} \right\}.
\end{equation}
\end{widetext}
In Eq. (\ref{eq:rhotpt_viscoelastic}), $\chi(t)$ is a time-dependent function that is defined through its Laplace transform as  
\begin{equation}\label{eq:chi}
	  \tilde{\chi}(s)  \equiv \int_0^{\infty} dt \, e^{-st} \chi(t) = \frac{{\eta}(s)}{s {\eta}(s) - \frac{\kappa}{6\pi a}}, 
\end{equation}
i.e., it is directly linked to the viscoelasticity of the fluid quantified by the frequency-dependent viscosity of Eq. (\ref{eq:etas}), and arises from the evolution equation (\ref{eq:GLE}).
It should be noted that $\tilde{\chi}(s)$ has at least one pole if $\kappa$ is non-zero. Hence, the late time behaviour of $\chi(t)$ will be dominated by the largest positive root  $s = \Omega > 0$ of the so-called Kramers-Grote-Hynes equation~\cite{grote1980,grote1981}, which takes the form 
\begin{equation}\label{eq:GroteHynes}
    \Omega {\eta}(\Omega) - \frac{\kappa}{6 \pi a} = 0.
\end{equation}
Note that the special frequency $\Omega$ corresponds to the reactive eigenvalue that gives rise to the unstable motion of the particle over the barrier. Therefore, the function $\chi(t)$, which is numerically computed by means of the Talbot method~\cite{talbot1979}, generally behaves at late time as $\chi(t)\sim  e^{\Omega t}$ at leading order.

It must be pointed out that Eq. (\ref{eq:rhotpt_viscoelastic}) is derived by applying open boundary conditions, hence the trajectories over the barrier can recross the points $x = \pm x_0$ in principle and do not exactly correspond to transition paths. However, this approximation for the probability density $\varrho(t_{\mathrm{TP}})$ becomes very good for sufficiently high barriers, since the particle escapes with a high probability to $x=\pm\infty$ once it leaves the transition region on one side, making recrossing events of the boundaries unlikely~\cite{zhang2007,pollak2016}. Indeed, in Figs. \ref{fig:Figure3}(a) and \ref{fig:Figure3}(b) we verify that the analytic approximation (\ref{eq:rhotpt_viscoelastic}) of $\varrho(t_{\mathrm{TP}})$, shown by thick solid lines, is in very good quantitative agreement for both fluids with the experimentally measured distributions of $t_{\mathrm{TP}}$, provided that $\Delta U > 2.5~k_B T$. To compute the theoretical curves, the numerical approximations of $\chi(t)$ calculated through the inverse Laplace transform of Eq. (\ref{eq:chi}) combined with Eq. (\ref{eq:etas}) and the values of $\kappa$, $x_0$ and the temperature $T$ for each trapped particle are used in Eq.~(\ref{eq:rhotpt_viscoelastic}) without any tuning parameter. Therefore, these results demonstrate that the barrier-crossing process of the particle in a viscoelastic fluid is well described by the generalized Langevin model (\ref{eq:GLE}). However, given the intricate expression of this distribution, it is difficult to assess how it is affected by the specific viscoelastic and barrier parameters. The transition path dynamics of the bead can be better understood by calculating the first moment of the distribution, as detailed in Section \ref{sect:MTPT}.

\section{Mean transition path time}\label{sect:MTPT}

We now analyze the average duration of the transition paths. 
Once more, our starting point is to compare the experimental data with the expression of the mean transition path time predicted by the purely viscous diffusive model of the particle dynamics over the parabolic barrier. In this case, for sufficiently high energy barriers, the mean transition path time is given by~\cite{chung2014,laleman2017}
\begin{equation}\label{eq:MTPTNewtonian}
    \langle t_{\mathrm{TP}} \rangle = \tau \ln \left( 2 e^{\gamma_{\mathrm{EM}}} \frac{\Delta U}{k_B T}\right),
\end{equation}
where $\gamma_{\mathrm{EM}} \approx 0.5772$ is the Euler-Mascheroni constant. Eq. (\ref{eq:MTPTNewtonian}) implies that, when rescaling $\langle t_{\mathrm{TP}} \rangle$ by the characteristic time-scale $\tau = \gamma/\kappa=6\pi a \eta_0/\kappa$ , the resulting ratio depends logarithmically on the barrier height only and is independent of the curvature and viscosity. 

In Fig. \ref{fig:Figure3}(c) we depict as a solid line the ratio $\langle t_{\mathrm{TP}} \rangle / \tau$ described by~Eq. (\ref{eq:MTPTNewtonian}). We also plot the experimental results for the corresponding ratio in the polymer solution for distinct energy barriers, where we use the zero-shear value $\eta_0 = 0.0093$ Pa s of this fluid in the calculation of $\tau$. Two important observations can be made regarding the behavior of $\langle t_{\mathrm{TP}} \rangle / \tau$ in the polymer solution. The first one is that all the experimental data points fall well below the curve predicted by Eq. (\ref{eq:MTPTNewtonian}). 
This is consistent with the fact that, as the particle surmounts the barrier, it deforms the surrounding polymer fluid, thus probing non-zero frequency components of its viscosity that are always smaller than $\eta_0$. As a result, the drag force is effectively reduced compared to the one that would be exerted in a Newtonian fluid of constant viscosity $\eta_0$, thus shortening the average duration of the transition paths. 

The second important property of the rescaled time $\langle t_{\mathrm{TP}} \rangle / \tau$ in the polymer solution is its nontrivial dependence with $\kappa$. This can be discerned in Fig. \ref{fig:Figure3}(c), where there is a significant dispersion among data points having similar values of $\Delta U$ but different values of $\kappa$. Such a dispersion is larger than the error bars, in stark contrast with the expected data collapse of $\langle t_{\mathrm{TP}} \rangle / \tau$ in a Newtonian fluid, which is described by the $\kappa$-independent logarithm in Eq. (\ref{eq:MTPTNewtonian}). Indeed, as demonstrated in Fig. \ref{fig:Figure3}(c), we check that the experimental values of $\langle t_{\mathrm{TP}} \rangle / \tau$ are in accord with Eq.~(\ref{eq:MTPTNewtonian}) in the Newtonian glycerol/water mixture at 55\%~wt, whose constant viscosity ($0.0076$~Pa~s) is very close to the zero-shear viscosity of the polymer solution. Furthermore, we numerically computed the mean transition path time of the bead in the viscoelastic fluid by using the analytic approximation of Eq. (\ref{eq:rhotpt_viscoelastic}) for $\varrho(t_{\mathrm{TP}})$ and the identity $\langle t_{\mathrm{TP}} \rangle = \int_0^{\infty} dt_{\mathrm{TP}}\,t_{\mathrm{TP}} \varrho(t_{\mathrm{TP}})$. Upon dividing by $\tau$, we obtain the shaded area depicted in Fig. \ref{fig:Figure3}(c). Here, the color-code reproduces the range of values of $\kappa$ explored in the experiments, i.e., $1.5~p\mathrm{N}\mu\mathrm{m}^{-1} < \kappa < 3.0~ p\mathrm{N}\mu\mathrm{m}^{-1}$. By increasing $\kappa$ at fixed $\Delta U$, the quantity $\langle t_{\mathrm{TP}} \rangle / \tau$ decreases. Remarkably, we find that the  scattered data fully lie inside this shaded area, thus verifying again that the model of Eq. (\ref{eq:GLE}) is able to quantitatively account for the viscoelastic effects of the medium on the transition paths of the particle over the barrier. 

Moreover, in Fig. \ref{fig:Figure3}(e) we represent as circles the dependence on $\kappa$ of the experimental rescaled times $\langle t_{\mathrm{TP}} \rangle / \tau$ in the polymer solution. These values are in good agreement with the curves obtained theoretically by varying $\kappa$ at fixed $\Delta U$. The latter are depicted as stripes in Fig. \ref{fig:Figure3}(d), where each color change represents an increment of $0.5k_B T$ in the barrier height $\Delta U$, which ranges from $2.5k_BT$ to $6k_BT$.
These findings reveal a new effect, namely, 
a monotonically decreasing behavior of $\langle t_{\mathrm{TP}} \rangle / \tau$ with increasing $\kappa$, which is entirely absent in the case of transition paths taking place in Newtonian fluids.

Similar effects to those described above are observed for the mean transition path time in the micellar solution, where $\langle t_{\mathrm{TP}} \rangle / \tau$ appears to be roughly one order of magnitude shorter than the Newtonian value of Eq. (\ref{eq:MTPTNewtonian}) [Fig. \ref{fig:Figure3}(d)], while the monotonic decrease with respect to the curvature of the barrier is clearly noticeable in Fig. \ref{fig:Figure3}(f). For the micellar solution, we used the value $\eta_0 = 0.023$ Pa s in the calculation of the time-scale $\tau$. Once again, by performing control experiments in a Newtonian water-glycerol mixture at 70\%~wt, whose viscosity ($0.021$~Pa~s) is similar to the above zero-shear viscosity, we confirm that the non-trivial behavior of $\langle t_{\mathrm{TP}} \rangle / \tau$ in the micellar solution can be attributed to its viscoelastic nature. These experimental values are very well described by numerical calculations using Eq. (\ref{eq:rhotpt_viscoelastic}), represented by the shaded area of Fig. \ref{fig:Figure3}(d).  Furthermore, it is interesting to note that in the micellar solution, the rescaled time displays a much sharper drop with respect to Eq. (\ref{eq:MTPTNewtonian}) than in the polymer solution for similar values of $\Delta U$ and $\kappa$, thus revealing an additional dependence on the specific viscoelastic spectrum of the fluid. As demonstrated in the following Section, all these effects originate from the selection by the particle of the reactive frequency $\Omega$ introduced in Section \ref{sect:TPTD}, which is predominant during successful barrier crossing events.

\section{Discussion}\label{sect:disc}

\begin{figure*}
\centering
 \includegraphics[width=0.975\textwidth]{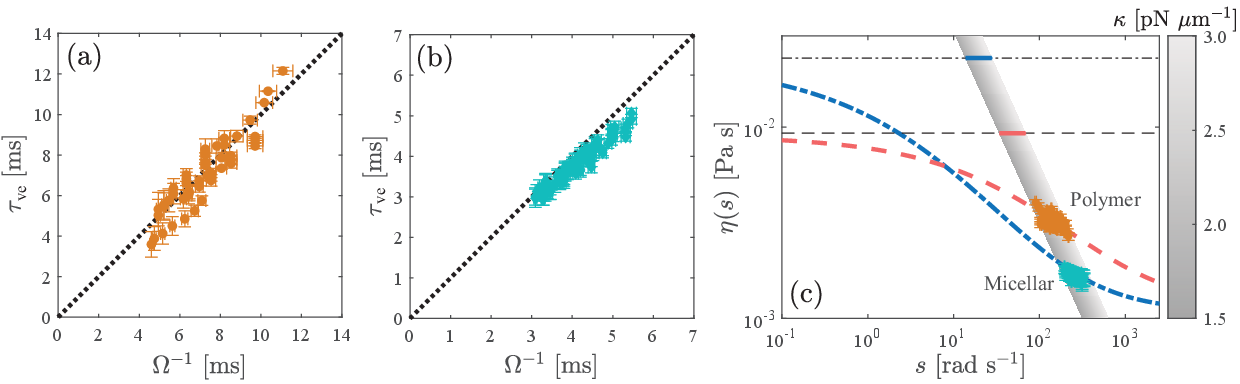}
 \caption{Comparison between the characteristic time-scale $\tau_{\mathrm{ve}}$ experimentally measured from the transition paths by use of Eq. (\ref{eq:tauk}) and the inverse of the reactive frequency $\Omega$ computed independently from Eq. (\ref{eq:GroteHynes}) using the linear microrheology measurements described by Eq. (\ref{eq:etas}) in the case of: (a) the polymer, and (b) the micellar solution. In (a) and (b), the identity line is represented as a dotted line. (c) Effective viscosity probed by the bead along transition paths in the polymer (orange $\circ$ symbols) and \textit{} micellar (cyan $\diamond$ symbols) solutions as a function of their  reactive frequency. The corresponding microrheological curves are depicted as a dashed (polymer solution) and a dotted-dashed line (micellar solution). The shaded area represents the  curves $\kappa/(6\pi a s)$ over the range $1.5\, \mathrm{pN}\,\mu\mathrm{m}^{-1} < \kappa < 3.0\, \mathrm{pN}\,\mu\mathrm{m}^{-1}$ spanned in the experiments, whose values are indicated by the grayscale. Horizontal dashed and dotted-dashed lines outline the  values of the zero-shear viscosity $\eta_0$ of the polymer and micellar solution, respectively, whose intersections with the shaded area (thick solid lines) determine shorter reactive frequencies $\Omega_0$ in the case of Newtonian fluids with the same viscosity as $\eta_0$.}
\label{fig:Figure4}
\end{figure*}

Since we have verified that Eq. (\ref{eq:rhotpt_viscoelastic}) describes well the statistics of the transition path time of the bead in the two surrounding viscoelastic fluids, we can further exploit this formula in light of the experimental data 
to elucidate the physical origin of the deviations from the Newtonian predictions. An important observation about the behavior of the theoretical values of $\langle t_{\mathrm{TP}} \rangle / \tau$ for both fluids is that, at constant $\kappa$, they follow an approximate linear relationship with $\ln [\Delta U / (k_B T)]$. This is clear from Figs. \ref{fig:Figure3}(c) and \ref{fig:Figure3}(d), 
where the upper and lower boundaries of the shaded areas corresponding to the extreme values of $\kappa$ explored in the experiments, follow straight lines. Consequently, this suggests that a phenomenological expression for the mean transition path time of the type of Eq.~(\ref{eq:MTPTNewtonian}) should also hold in a viscoelastic fluid, where  $\tau = \gamma / \kappa$ should be replaced by a more general time-scale 
$\tau_{\mathrm{ve}}$ that depends non-linearly on $\kappa^{-1}$ as well as on the viscoelastic parameters of the fluid, but not on the barrier height. As a matter of fact, an expression for the mean transition path time with a logarithmic dependence on $\Delta U$ have been formally derived for generalized Langevin models with 
power-law~\cite{carlon2018} and exponential memory kernels~\cite{medina2018}. 
In these models, the time-scale $\tau_{\mathrm{ve}}$ is exactly given by the inverse of the reactive frequency $\Omega$ that results from the solution of the corresponding Kramers-Grote-Hynes equation~(\ref{eq:GroteHynes}). For the experimental system studied here, it is not easy to derive an analytic expression for $\langle t_{\mathrm{TP}} \rangle$ from the density
$\varrho(t_{\mathrm{TP}})$ due to the more complex behavior of the memory kernel or frequency-dependent viscosity in Eq. (\ref{eq:etas}). Nevertheless, the reactive frequency $\Omega$ can be obtained in a straightforward manner from numerically solving  Eq. (\ref{eq:GroteHynes}) for each experimental value of $\kappa$. Therefore, we can test directly from the experimental data whether the time-scale $\tau_{\mathrm{ve}}$ introduced above is equal to $\Omega^{-1}$ in our experiments. To this end, for each transition region we estimate $\tau_{\mathrm{ve}}$ through the ratio 
\begin{equation}\label{eq:tauk}
    \tau_{\mathrm{ve}} = \frac{\langle t_{\mathrm{TP}} \rangle} { \ln \left( 2 e^{\gamma_{\mathrm{EM}}} \frac{\Delta U} {k_B T} \right)}, 
\end{equation}
and compare it with the corresponding value of $\Omega^{-1}$. The results of this comparison for the polymer and micellar fluids are presented in Figs. \ref{fig:Figure4}(a) and \ref{fig:Figure4}(b), respectively, where we verify experimentally in both cases that the relation $\tau_{\mathrm{ve}} \approx \Omega^{-1}$ holds remarkably well. Therefore, by virtue of Eq. (\ref{eq:GroteHynes}), $\tau_{\mathrm{ve}}$ can be written as
\begin{equation}\label{eq:taukappa}
    \tau_{\mathrm{ve}} \approx \frac{6\pi a \eta(\Omega)} {\kappa},
\end{equation}
which allows us to recast the mean transition path time of the bead in a viscoelastic fluid as
\begin{equation}\label{eq:MTPTgeneral}
    \langle t_{\mathrm{TP}} \rangle \approx \frac{6\pi a \eta(\Omega)}{\kappa} \ln \left( 2 e^{\gamma_{\mathrm{EM}}} \frac{\Delta U}{k_B T}\right).
\end{equation}
Noteworthy, Eq. (\ref{eq:MTPTgeneral}) is consistent with the fact that only those events for which the reactive frequency $\Omega$ is thermally activated give rise to transition paths, along which the particle  behaves as effectively moving in a fluid of constant viscosity  $\eta(\Omega)$ that lies in the interval  $\eta_{\infty} < \eta(\Omega) < \eta_0$. In Fig. \ref{fig:Figure4}(c) we represent as a function of $\Omega$ the experimental values of this effective viscosity probed by the particle along transition paths, where $\eta(\Omega)$ is deduced from the measured $\tau_{\mathrm{ve}}$ using Eq. (\ref{eq:taukappa}). As expected, the effective viscosity agrees well with the microrheological curve of each fluid at the corresponding frequencies. This plot clearly reveals why the reduction of  $\langle t_{\mathrm{TP}} \rangle / \tau$ is more pronounced in the micellar solution than in the polymer one, despite of the fact that the zero-shear viscosity of the former is larger than that of the latter. As depicted in Fig. \ref{fig:Figure4}(c) for a fixed value of the curvature of the barrier, the curve $\kappa/(6\pi a s)$ intersects the microrheological curve $\eta(s)$ of the micellar solution at a higher reactive frequency than the polymeric one. The corresponding value of $\eta(\Omega)$ is also smaller than that of the polymer solution. In both viscoelastic fluids, the reactive frequencies $\Omega$ are always higher than their Newtonian counterparts of same zero-shear viscosity $\eta_0$, i.e., $\Omega_0 \equiv \kappa / (6 \pi a \eta_0) = \tau^{-1}$, thus implying that $\tau > \tau_{\mathrm{ve}}$ according to Eqs. (\ref{eq:GroteHynes}) and (\ref{eq:taukappa}). This clearly explains the observed reduction of $\langle t_{\mathrm{TP}} \rangle / \tau$ in Figs. \ref{fig:Figure3}(c) and (d). Likewise, the decrease of $\langle t_{\mathrm{TP}} \rangle / \tau$ with $\kappa$ at fixed $\Delta U$ in Figs. \ref{fig:Figure3}(e) and \ref{fig:Figure3}(f) is a consequence of the monotonically decreasing behavior of $\tau_{\mathrm{ve}}/\tau = \eta(\Omega) / \eta_0$ with $\Omega$ as $\kappa$ increases. Therefore, these effects are direct results of the linear viscoelasticity of the fluids used as non-Markovian baths, where a higher degree of elasticity translates into a more pronounced drop of $\eta(s)$ with frequency.

\section{Conclusion}\label{sect:conc}

We have experimentally investigated  the statistics of the duration of transition paths of a colloidal bead traversing an energy barrier between two optical potential wells in a viscoelastic fluid. We have verified that the measured transition path time distributions can be quantitatively described by an expression derived from a generalized Langevin model that includes the viscoelasticity of the medium. Therefore, we have confirmed for the first time the validity of this theoretical approach for describing in detail the barrier-crossing dynamics of colloidal particles in multistable potentials, beyond the paradigm of Markovian baths. By analyzing the average duration of transition paths for distinct bistable potentials, we have been able to identify a logarithmic dependence of this quantity on the barrier height and its proportionality to a time-scale determined by the barrier curvature and the viscoelasticity of the medium. On this basis, we have experimentally verified that such a time-scale coincides with the inverse of the frequency of the reactive mode, thereby establishing the frequencies at which the viscosity of fluid is effectively probed by the particle during the successful escape events over the barrier.

\section*{Acknowledgments}
We acknowledge support from DGAPA-UNAM PAPIIT Grants No. IA104922, IN104924 and PIIF-2-24.

\newpage

\appendix

\section{Derivation of the transition path distribution in a non-Markovian bath}\label{app:derivTPTP}

To find an approximate formula for the transition path time distribution of the bead in a viscoelastic fluid for sufficiently high parabolic barriers, we can follow the approach described in \cite{laleman2017,singh2021}. The derivation relies on the absorption function $Q(t)$, which gives the probability that the particle has
already crossed the boundary at $+x_0$ at time $t > 0$ at least once, starting at time $t=0$ from $-x_0$. This can be expressed in terms of the conditional probability density of finding the particle at position $x$ at time $t>0$ given that it was located at position $-x_0$ at time $t=0$, denoted as $P(x,t|-x_0,0)$. The absorption function is 
\begin{equation}\label{eq:absorp}
    Q(t) = \int_{+x_0}^{\infty} dx \,  P(x,t|-x_0,0).
\end{equation}
Since the difference $Q(t+dt) - Q(t)$ represents the fraction of particle trajectories that cross the point $+x_0$ during the time interval $[t,t+dt]$, then 
the normalized probability density function of the transition path time $t_{\mathrm{TP}}$ is given by
\begin{equation}\label{eq:rhoTPTabsorb}
    \varrho(t_{\mathrm{TP}}) = \frac{1}{Q(\infty)} \left.\frac{dQ(t)}{dt}\right|_{t = t_{\mathrm{TP}}},
\end{equation}
where the term $Q(\infty) = \lim_{t \rightarrow \infty} Q(t)$ arises from the normalization of $\varrho(t_{\mathrm{TP}})$ and the fact that $Q(0) = 0$ according to Eq. (\ref{eq:absorp}). As discussed in \cite{laleman2017}, Eq. (\ref{eq:rhoTPTabsorb}) is just an approximation because it does not impose the correct absorbing boundary conditions at $\pm x_0$ that are required in the definition of a transition path, thus also counting trajectories that can recross these two points. Nevertheless, if the barrier is sufficiently high such recrossings become extremely uncommon as the particle escape to $x=-\infty$ (when it does not succeed in crossing the barrier) or $x=+\infty$ (when it succeeds). Hence, Eq. (\ref{eq:rhoTPTabsorb}) represents an excellent approximation for the actual transition path time distribution, where $P(x,t|-x_0,0)$ is the conditional probability with open boundary conditions.

As shown in \cite{okuyama1986}, a generalized Smoluchowski equation corresponding to the generalized Langevin equation (\ref{eq:GLE}) for $P(x,t|-x_0,0)$ with open boundary conditions, can be written as
\begin{widetext}
\begin{equation}\label{eq:genSmol}
    \frac{\partial P(x,t|-x_0,0)}{\partial t}
     = -\frac{d \ln \chi(t)}{dt} \left\{ \frac{\partial}{\partial x} \left[x P(x,t|-x_0,0)\right] + \frac{k_B T}{\kappa} \frac{\partial^2 P(x,t|-x_0,0)}{\partial x^2}\right\}.
\end{equation}
\end{widetext}
In~(\ref{eq:genSmol}), $\chi(t)$ is a dimensionless function dependent on the elapsed time $t \ge 0$, which is defined in Eq. (\ref{eq:chi}) and generally diverges as $t \rightarrow \infty$ due to the positive pole of its inverse Laplace transform.

Eq.~(\ref{eq:genSmol}) can be readily solved in the Fourier domain and then $P(x,t|x_0,0)$ can be found by Fourier inversion of the wave-number dependent solution with the initial condition $P(x,0|x_0,0) = \delta (x-x_0)$ and open boundary conditions. Indeed, by expressing $P(x,t|x_0,0)$ for a fixed value $x_0$ in terms of its spatial Fourier transform, $\hat{P}(k,t|x_0,0)$, 
\begin{equation}\label{eq:FourP}
    P(x,t|x_0,0) = \frac{1}{2\pi} \int_{-\infty}^{\infty} dk \, e^{ikx} \hat{P}(k,t|x_0,0),
\end{equation}
in the Fourier domain, Eq.(\ref{eq:genSmol}) becomes
\begin{widetext}
\begin{equation}\label{eq:FourierFP}
    \frac{\partial \hat{P}(k,t|-x_0,0)}{\partial t} = \frac{d \ln \chi(t)}{dt} k \frac{\partial \hat{P}(k,t|-x_0,0)}{\partial k} + \frac{k_B T}{\kappa}\frac{d \ln \chi(t)}{dt} k^2 \hat{P}(k,t|-x_0,0),
\end{equation}
\end{widetext}
where the Fourier transform of the initial condition is $\hat{P}(k,0|x_0,0) = e^{+ikx_0}$. Then,~Eq.(\ref{eq:FourierFP}) can be solved by, e.g., the method of characteristics, thus obtaining
\begin{equation}
     \hat{P}(k,t|x_0,0) =  \exp \left\{ +ikx_0 \chi(t) -\frac{k_B T}{2 \kappa}  \left[\chi(t)^2 -1\right] k^2 \right\},
\end{equation}
whose inverse Fourier transform is
\begin{widetext}
\begin{equation}\label{eq:solSmol}
    P(x,t|-x_0,0) = \sqrt{\frac{\kappa}{2\pi k_B T\left[\chi(t)^2 - 1\right]}} \exp \left\{ -\frac{\kappa}{2k_BT} \frac{\left[ x + x_0 \chi(t) \right]^2}{ \chi(t)^2-1}\right\},
\end{equation}
\end{widetext}
and corresponds to a Gaussian function with mean $-x_0 \chi(t)$ and variance $k_B T [\chi(t)^2 -1]/\kappa$. Therefore, by inserting the expression of Eq. (\ref{eq:solSmol}) into Eq.~(\ref{eq:absorp}), the absorption function takes the form
\begin{equation}\label{eq:absorption}
    Q(t) = \frac{1}{2} \mathrm{erfc}\left( \sqrt{\frac{\kappa x_0^2}{2k_B T}}\sqrt{\frac{\chi(t)+1}{\chi(t)-1}} \right),
\end{equation}
which implies that 
\begin{equation}\label{eq:infabs}
    Q(\infty) = \frac{1}{2} \mathrm{erfc}\left( \sqrt{\frac{\kappa x_0^2}{2k_B T}} \right),
\end{equation}
because of the divergence of $\chi(t)$ as $t \rightarrow \infty$. Finally, by substituting  Eqs. (\ref{eq:absorption}) and (\ref{eq:infabs}) into Eq. (\ref{eq:rhoTPTabsorb}), we obtain the expression of the transition path time distribution given by Eq. (\ref{eq:rhotpt_viscoelastic}).


\end{document}